\documentclass[aps,pra,twocolumn,amsmath,amssymb]{revtex4-2}
\usepackage{graphicx}
\usepackage[colorlinks=true,citecolor=blue,linkcolor=blue,urlcolor=blue]{hyperref}
\usepackage{amsmath,amssymb,bm,physics}
\usepackage[colorlinks=true,citecolor=blue,linkcolor=blue,urlcolor=blue]{hyperref}
\usepackage{comment}
\usepackage[caption=false]{subfig}

\begin{document}

\title{Fermionic Stoner-Dicke phase transition in Circuit Quantum Magnetostatics}

\author{Adel Ali}
\affiliation{Department of Physics and Astronomy, Texas A\&M University, College Station, TX, 77843 USA}
\author{Alexey Belyanin}
\affiliation{Department of Physics and Astronomy, Texas A\&M University, College Station, TX, 77843 USA}

\begin{abstract}
We present a minimal tunable many-body system of fermions coupled to quantum magnetic flux, which is analytically diagonalizable and exhibits a variety of many-body phenomena such as Stoner orbital instability and Dicke-like quantum phase transition. In contrast to standard cavity quantum electrodynamics with its electric-dipole coupling of the electric field operators with matter, here it is the quantized magnetic field of an LC-resonator which is coupled to the angular momentum of particles. Adding the Josephson junction (JJ) to the linear LC circuit allows us to explore nonlinear flux-matter phases and sector-selective photon dressing in regimes relevant to circuit QED and mesoscopic rings. Furthermore, we consider the tight-binding systems that exhibit a tunable nonlinearity representing artificial JJ, but without actual JJs included in the circuit.  

%a way to engineer a many-body QPT in a system that is simple enough to write down analytically, but rich enough to be experimentally relevant. stoner orbital instability and Dicke like phase transition are found in a model of fermions on a ring coupled to quantum magnetic flux (QMF). 

%We couple  magnetic field with charge in minimal coupling. No RWA , keeping diamagnetic $A^2$. Since the single mode here is coupled to fermions the no-go theorem is not applicable and one can get a fermionic superradiance first-order phase transition without approximation and keeping gauge invariance by not dropping the diamagnetic $A^2$ term. 

 %For spinless fermions , we compare the balanced configuration (minimal $\sum m_i^2$ at $M\!=\!\sum m_i=0$) against the fully polarized configuration ($m=0,1,\dots,N-1$), and obtain a sharp level crossing for even $N\ge4$ at a finite critical coupling, while for odd $N$ the crossing occurs only in a singular limit. All formulas are closed form.  
%No JJ required (so higher Q factor) , the system may operate in higher temperatures as C and L are tunable. We also show that for a tigth binding ring coupled to the resonator the exact peirls substiution will give rise to effictive tunable JJ for the resonator. 

\end{abstract}

\maketitle

Strong light-matter interaction is at the heart of cavity quantum electrodynamics (CQED) which describes the coupling of electric field operators of quantized cavity modes with matter. Various  CQED models at strong coupling have revealed routes to engineer interactions, ordering, and transport far beyond bare Coulomb forces~\cite{Blais2021,Schlawin2019}.

In conventional atomic CQED, light--matter coupling is overwhelmingly governed by the
electric-dipole interaction, while magnetic-dipole couplings are relativistically suppressed, rendering them typically orders of magnitude weaker than their
electric-dipole counterparts \cite{Walther2006}.
In mesoscopic platforms, however, the relevant magnetic moment is often orbital: circulating
currents in quantum rings or quantum Hall edge channels generate an effective dipole
\(\mu_{\rm orb}\sim I A\), which can greatly exceed the Bohr magneton and thereby enhance coupling
to the magnetic near field of microwave photons \cite{Buttiker1983,Levy1990,B...Jayich2009,Bosco2019PRB}.

In mesoscopic and synthetic systems, \emph{vector potentials themselves}, even in nominally field-free regions, can control quantum phases, as exemplified by the Aharonov--Bohm (AB) effect and its modern realizations in solid-state and cold-atom platforms~\cite{Dalibard2011,Goldman2014,PhysRevA.110.022604}. This offers a topological protection of the coupling inherited from the AB effect \cite{PhysRevA.110.022604}. A central challenge is to formulate gauge-invariant, analytically tractable models where a \emph{quantized} gauge field mediates effective, tunable interactions among mobile particles, while avoiding pathologies associated with superradiant instabilities and no-go constraints tied to the diamagnetic term~\cite{HeppLieb1973,Rzazewski1975,NatafCiuti2010,Viehmann2011,Kim2025}.

%in \cite{Kim2025} the authors avoided the no-go theorem by replacing the photon mode with magnon mode (spins) where the exchange coupling wont naturally have the diamagnetic term. Here the full minimal coupling Hamiltonian is considered beyond the dipole approximatim

We address this challenge by constructing  exactly diagonalizable coupled flux-matter models within the domain of \emph{cavity quantum magnetostatics} (CQM) where a quantized transverse magnetic near field, generated, e.g., by a supercurrent in a superconducting loop, couples minimally to mobile charges on a quantum ring (QR) through their orbital motion; see Fig.~1.  

The CQM formulation differs qualitatively from standard CQED: the leading interaction is magnetic rather than electric dipole, selection rules are determined by the angular momentum rather than polarization charge, and collective effects arise from the \emph{angular momentum channel} generated by the quantized flux.

Ideally, a single superconducting loop is an inductor. We will ignore any small parasitic resistance due to external wiring in this proof of concept treatment. The capacitance of a loop is also small, but it can be introduced on purpose to tune the resonant frequency. Therefore, 
we model a single superconducting loop as a lumped linear LC circuit with canonical flux-charge pair \((\hat\Phi,\hat Q)\) obeying \([\hat\Phi,\hat Q]=i\hbar\). Quantization yields the harmonic-oscillator Hamiltonian
\begin{equation}
\hat H=\frac{\hat Q^{2}}{2C}+\frac{\hat\Phi^{2}}{2L}
=\hbar\omega\!\left(\hat a^\dagger\hat a+\tfrac{1}{2}\right), 
\quad \omega=\frac{1}{\sqrt{LC}},
\label{eq:H-LC}
\end{equation}
with ladder-operator representation
\begin{equation}
\hat\Phi=\Phi_{\mathrm{zpf}}\left(\hat a+\hat a^\dagger\right),\quad
\hat Q=i\,Q_{\mathrm{zpf}}\left(\hat a^\dagger-\hat a\right),
\label{eq:zpf}
\end{equation}
where zero-point amplitudes are 
$\Phi_{\mathrm{zpf}}=\sqrt{\frac{\hbar Z}{2}},\ \ 
Q_{\mathrm{zpf}}=\sqrt{\frac{\hbar}{2Z}}$
and \(Z=\sqrt{L/C}\) is the characteristic impedance.

%%%%%%%%%%%%%%%%%%%%%%%%%
\begin{figure}[t]
\includegraphics[width=1\linewidth]{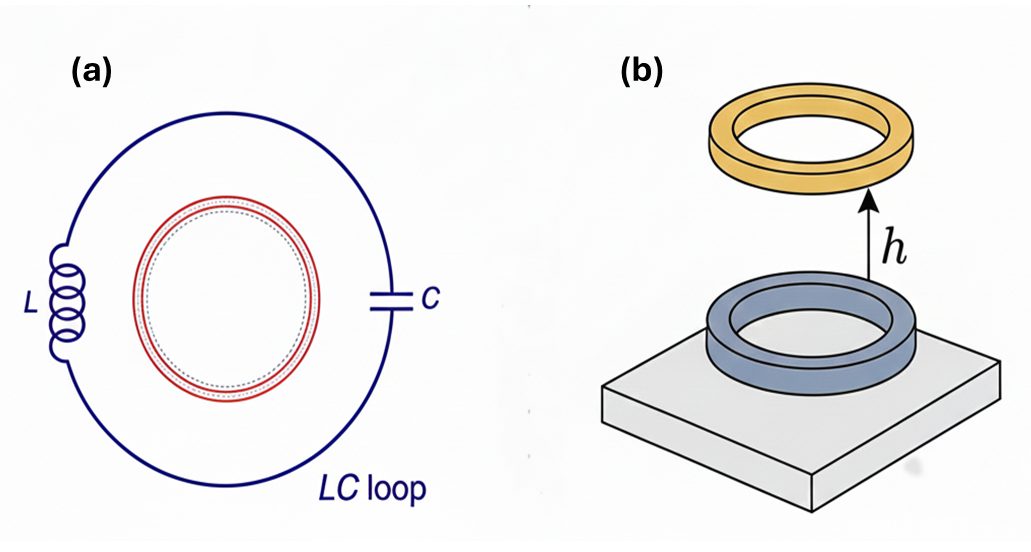}
    \caption{Sketches of a coupled QR--superconducting LC loop system. (a) Top view with the QR (red circle) inside the LC loop.  (b) Side view: the QR above can be positioned at variable distance $h$ from a superconducting loop, e.g., by using a spacer layer, to control coupling. }
    \label{fig:qr-lc}
\end{figure}

In the Coulomb gauge, 
%with $\boldsymbol{\nabla}\!\cdot\!\hat{\mathbf A}=0$ and $\hat{\mathbf B}=\boldsymbol{\nabla}\!\times\!\hat{\mathbf A}$, 
the quantized vector potential generated by the supercurrent of a single mode can be written as
\begin{equation}
\hat{\mathbf A}(\mathbf r)=\Phi_{\rm zpf}\,\mathbf f(\mathbf r)\,\big(\hat a+\hat a^\dagger\big),
\qquad [\hat a,\hat a^\dagger]=1. 
\label{eq:A-field}
\end{equation}
 The function $\mathbf f(\mathbf r)$ encodes the magnetostatic mode profile set by the loop/inductor geometry. We assume that by design the near electric field at the QR is negligible. 

The LC circuit here has a passive role of a cavity, in a sense that it does not need to be controlled or externally excited like a qubit. Therefore, there is freedom in tuning $L$ and $C$ towards higher resonant frequencies $\omega$, which could be important for increasing the operating temperatures, for example, when discussing the possibility of cavity-induced superconductivity. In addition to that, $LC$ cavities can be coupled inductively or capacitively to enhance or suppress certain interaction.   One can refer to the coupling as electron-fluxon. So, here the cavity enhancement means that the magnetic field is strengthened by having a low inductance $L$. 
%makes the upper limit is the SC gap which can reach Thz for some materials. The effect of the cavity loop is derived from the zero point fluctuations in the ground state which has zero average flux. 

The Dicke superradiant phase transition (SPT) has been realized mainly in a driven–dissipative (i.e., pumped and open) Dicke-model \cite{openDicke}, while its observation in a genuine equilibrium situation remains elusive. It was argued that the no-go theorems for the SPT in CQED persist in the context of superconducting circuit QED \cite{Viehmann2011}.  Our work aims to provide a platform that avoids these no-go theorems and enables access to the SPT in the equilibrium regime.

%The excitations of the LC circuit are usually called photons. However, here we focus on the near magnetic field generated by the current coupling to magnetic moments of matter. 

%Achieving Dicke phase transition has been done in non-equilibrium pumped and open Dicke models settings, in equilibrium it is still an elusive goal.

%and a small size loop representing the cavity, quantum because the LC loop is quantized and magnetostatic as the focus is on the near magnetic field generated by the current coupling to magnetic moments of matter.

%Looking at the ground state of the LC, the flux variable equivalently the current is gaussian centered around zero. 
%%%%%%%%%%%%%%%%%%%%%%%%%%%%%%%%%%%%%%%%%%%%%%%%%%%%%%%%%%%%%%%%%%%%%%%%%%%
\section{Electrons on a ring} 

For $N$ charges of mass $m$ and charge $q$ on a QR of radius $R$, we adopt the gauge-invariant minimal-coupling Hamiltonian
\begin{equation}
\hat H=\sum_{i=1}^N\frac{1}{2m_0}\!\left[\hat{\mathbf p}_i-q\,\hat{\mathbf A}(\hat{\mathbf r}_i)\right]^{\!2}
+ U_c(\hat r_i) + U_{QR}(\hat r_i)+\hbar\omega\,\hat a^\dagger\hat a,
\label{eq:H-magnetostatic}
\end{equation}
without rotating-wave approximation (RWA) and   retaining the diamagnetic $A^2$ term. Zeeman coupling to spins will be included in the next section.  Here $m_0$ is free electron mass and $U_c(\hat r_i)$ and $U_{QR}(\hat r_i)$ are the potential of the crystal lattice and the confining ring potential. The part $\frac{\hat{\mathbf p}_i^2}{2m_0} + U_c(\hat r_i) $ gives rise to energy bands. In this section we assume low-energy electrons in a parabolic conduction band with effective mass $m_{\rm eff}$, i.e., the kinetic energy $\frac{\hat{\mathbf p}_i^2}{2m_{\rm eff}} $. Furthermore, the confining ring potential forces the electrons to move around the ring with the angular coordinates $\theta_i$, and we can replace linear momenta with angular momenta  $\hat L_i = -i \hbar \frac{\partial}{\partial \theta_i}$.   
% The potential $U$ reshapes the single-particle spectrum; reshaping can also come from moire graphene where the bands are tunable. Projecting to a low-energy band near its minimum, the dispersion is generically quadratic,
% $\varepsilon(m)\simeq \varepsilon_0+g_{\rm eff}(m-m_0)^2$,
%which motivates replacing $\hat H_{\rm mat}\to g_{\rm eff}\sum_i\hat L_i^2$ (up to an additive constant). $\hat M\equiv\sum_i\hat L_i$ 
This yields the model 
%while \emph{retaining} the microscopic $g$ in $\hat H_{\rm int}$
\begin{equation}
\hat H_{\rm low}
=
g_{\rm eff}\sum_{i=1}^N \hat L_i^2
+\hbar\omega\,\hat a^\dagger\hat a
+gN\phi^2\hat X^2
-2g\phi \hat X\hat M,
\label{eq:Hlow}
\end{equation}
where $g_{\rm eff} = \hbar^2/(2m_{\rm eff} R^2)$, $g=\hbar^2/(2m_0 R^2)$, $\hat M = \sum_i\hat L_i$, $\hat X\equiv \hat a+\hat a^\dagger$, and $\phi$ is the magnetic flux enclosed by the ring. 

For fixed $\hat M$ the bosonic sector of Eq.~\eqref{eq:Hlow} is a displaced and squeezed harmonic oscillator with renormalized frequency
\begin{equation}
\Omega=\sqrt{\omega\Big(\omega+\frac{4gN\phi^2}{\hbar}\Big)}.
\label{eq:Omega}
\end{equation}
Diagonalizing the cavity mode yields
\begin{equation}
\hat H_{\rm low}
=
g_{\rm eff}\sum_{i=1}^N \hat L_i^2
-\chi(\phi)\,\hat M^2
+\hbar\Omega\,\hat b^\dagger\hat b
+\frac{\hbar\Omega-\hbar\omega}{2},
\label{eq:HeffSep}
\end{equation}
where $\hat b$ is the normal-mode boson and the induced collective interaction is
\begin{equation}
\chi(\phi)=\frac{4g^2\phi^2}{\hbar\omega+4gN\phi^2}.
\label{eq:chi}
\end{equation}

The diagonalizing unitary transformation is equivalent to the single-mode squeeze operator
\begin{equation}
S(r)=\exp\!\Bigl[\frac{r}{2}\bigl(\hat a^{2}-\hat a^{\dagger 2}\bigr)\Bigr],\quad
\hat b=S^\dagger(r)\,\hat a\,S(r),
\nonumber 
\end{equation}
which acts on the canonically conjugate quadratures as
\begin{equation}
S^\dagger(r)\,\hat x\,S(r)=e^{-r}\hat x,\quad
S^\dagger(r)\,\hat p\,S(r)=e^{+r}\hat p.
\nonumber 
\end{equation}

The details of the diagonalization procedure are described in the Appendix. The eigenstates are displaced-squeezed harmonic oscillator states tensored with $\ket{\{m_i\}}$, where the quantum number $m_i$ is the eigenvalue of $\hat L_i/\hbar$ in the basis of angular momentum states. For the ground state of the system, Eq.~(\ref{eq:HeffSep}) shows competition between the  kinetic energy term $g_{\rm eff}\sum_i m_i^2$, which prefers a balanced (time-reversal-symmetric) distribution, and the cavity-induced attraction $-\chi M^2$ with $M=\sum_i m_i$, which prefers orbital polarization, as sketched in Fig.~2. This leads to the two possibilities for the ground state. 
%Thus, in the $\hat b$-vacuum,
%\begin{equation}
%\langle \hat x^2\rangle=\tfrac12 e^{-2r},\quad
%\langle \hat p^2\rangle=\tfrac12 e^{+2r},\quad
%\langle \hat x^2\rangle\,\langle \hat p^2\rangle=\tfrac14,
%\nonumber 
%\end{equation}

%In the cavity ground state $\hat b=0$ the many-body energetics reduces to
%\begin{equation}
%E(\{m_i\})
%=
%g_{\rm eff}\sum_{i=1}^N m_i^2
%-\chi(\phi)\,M^2,
%\qquad
%M=\sum_{i=1}^N m_i,
%\label{eq:Ematter}
%\end{equation}

For $N=2K+1$ the fermionic balanced configuration minimizing $\sum_i m_i^2$ is
$m=-K,\ldots,0,\ldots,K$, giving $M_{\rm bal}=0$ and
\begin{equation}
W_{\rm bal}\equiv\sum_{m=-K}^{K} m^2=\frac{K(K+1)(2K+1)}{3}.
\label{eq:Sbal}
\end{equation}
A competing polarized configuration (set by the band edge in the projected manifold) has total angular momentum $M_{\rm pol}\neq 0$ and kinetic weight $W_{\rm pol}=\sum_i m_i^2$. The balanced--polarized transition is a first-order level crossing determined by
\begin{equation}
\chi_c
=
g_{\rm eff}\,
\frac{W_{\rm pol}-W_{\rm bal}}{M_{\rm pol}^2-M_{\rm bal}^2}
=
g_{\rm eff}\,
\frac{W_{\rm pol}-W_{\rm bal}}{M_{\rm pol}^2}.
\label{eq:chic}
\end{equation}
Using Eq.~(\eqref{eq:chi}), the corresponding critical flux coupling value satisfies
\begin{equation}
\phi_c = \sqrt{\frac{g_{\text{eff}}\hbar\omega}{4gN (g - g_{\text{eff}})}}.
\label{eq:phi_c}
\end{equation}

\begin{figure}
    \centering
    \includegraphics[width=0.7\linewidth]{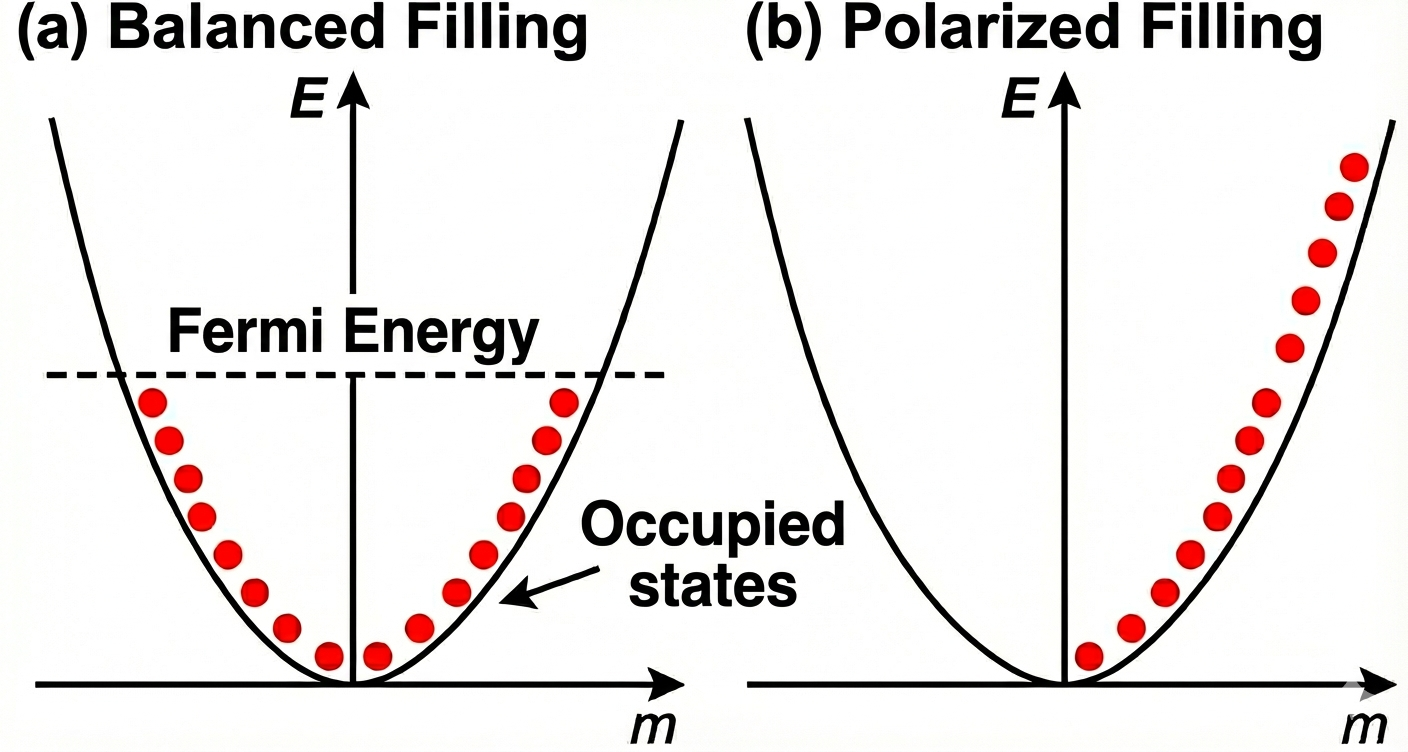}
    \caption{Schematic of the ground state filling for (a) balanced
  and (b) polarized electron configurations.}
    \label{fig:2}
\end{figure}

For $g_{\rm eff}$ to be smaller than  $g$ one needs an effective mass larger than the free electron mass. 
The model thus realizes a flux-driven \emph{orbital ferromagnet}, conceptually akin to Lipkin--Meshkov--Glick-type collective couplings~\cite{Lipkin1965} yet derived here microscopically from gauge-invariant minimal coupling, and distinct from a classic Stoner ferromagnetism~\cite{Stoner1938}. As we show below, an exact order parameter emerges: the cavity displacement $\langle \hat a+\hat a^\dagger\rangle\propto \langle \hat M\rangle$, with finite fluctuations across the jump.

The LC resonator frequency $1/\sqrt{LC}$  is tunable by changing $C$ and $L$. It can range from MHz up to THz.  The flux $\phi$ enclosed by the ring is geometry dependent, i.e., if the ring is off the plane of the resonator the flux will depend on the distance between them; see Fig.~1(b). 

The state of the flux variable can be measured with a coplanar SQUID inductively coupled to the resonator. 
%%%%%%%%%%%%%%%%%%%%%%%%%%%%%%%%%%%\subsection{Effective low-energy model and balanced--polarized transition (odd $N$)}

%%%%%%%%%%%%%%%%%%%%%%%%%%%%%%%%
\section{Zeeman coupling to electron spins} 

If the ring is in the region of a nonzero magnetic field of the cavity (as opposed to pure AB interaction with a quantized vector potential), then the spin of the electrons will couple to the LC magnetic field. In this case the cavity will mediate collective spin ferromagnetic interaction and collective spin-orbit coupling ``locking''. This can be used to engineer spin-orbit coupling in the materials which are naturally lacking it. The energies of the Hamiltonian 
\begin{equation}
\hat H= g_{\rm eff}\sum_{i=1}^N \hat L_i^{\,2}
+\hbar\omega\,\hat a^\dagger\hat a
+ gN\phi^2 \hat X^2
-\hat X\Big(2g\phi\,\hat M+\eta\,\hat\Sigma_z\Big)
\nonumber 
\label{eq:Hspin}
\end{equation}
are given by 
\begin{equation}
E_{n,\{m_i,s_i\}}
=
g\sum_{i=1}^N m_i^2
-\frac{\big(2g\phi\,M+\eta\,\Sigma\big)^2}{\hbar\omega+4gN\phi^2}
+\hbar\Omega\Big(n+\tfrac12\Big),
\nonumber 
\label{eq:spectrum}
\end{equation}
where $\eta \equiv \frac{g_s\mu_B}{2}B_{\rm zpf}$, $\Sigma$ is total spin, $g_s$ is the electron $g$-factor, $\mu_B$ the Bohr magneton, and $B_{\rm zpf}$ 
denotes the (zero-point) magnetic field amplitude at the ring.  
The same type of balanced-polarized transition is possible in this situation, but now it can be enhanced by spin polarization. As derived in the Appendix, the critical coupling for the transition described by the composite order parameter vector $\boldsymbol{\psi} = (M, \Sigma)^T$ is given by  
\begin{equation}
    \phi_c = \frac{1}{N} \sqrt{ \frac{\eta^2 - \frac{1}{4}N g_{\rm eff} \hbar \omega}{g(g_{\rm eff} - g)} }.
    \label{eq:critical_phi}
\end{equation}
This reduces to $\eta_c = \sqrt{\frac{g N \hbar \omega}{4}}$ when $g_{\rm eff}=g$ and the transition is purely spin driven. Note that the spin-driven transition is possible when $g_{\rm eff} > g$, i.e., the effective mass is smaller than free electron mass, in contrast to the orbital coupling driven transition.  
%This is similar to Stoner instability which is density dependent.

%%%%%%%%%%%%%%%%%%%%%%%%%%%%%%%%%%%%%%%%%%%%%%%%%%%%%%%%%%%%%%%%%%%%%

\section{Graphene ring coupled to quantized flux} 

While SPTs in our system can be realized for a variety of electron dispersions, we consider graphene quantum rings (GQRs) as another specific example,  which offers a minimal setting for Dirac carriers to acquire a controllable AB phase and thereby convert magnetic flux into a direct spectroscopic and transport knob. Recent gate-defined interferometers in high-mobility, encapsulated bilayer graphene exhibit clear multi-harmonic AB oscillations and long phase-coherence lengths \cite{Iwakiri2022NanoLett}, establishing GQRs as a clean platform for ballistic electron interferometry. In the quantum Hall regime, split-gate bilayer-graphene Fabry--P\'erot interferometers display pronounced AB oscillations of edge states \cite{Fu2023NanoLett}. Ring interferometry has also entered the moir\'e-correlated landscape: a gate-tunable interferometer in magic-angle twisted bilayer graphene reports flux-periodic oscillations \cite{Iwakiri2024NatCommun}. Complementary continuum and tight-binding analyses of bilayer and twisted-bilayer graphene rings map how confinement, stacking, and perpendicular field reshape the ring spectrum \cite{Mirzakhani2022PRB,Bandeira2025PRB}.

%$\hat\Phi=\Phi_{\rm zpf}(\hat a+\hat a^\dagger)$, i.e. the dimensionless flux threading the ring
%\begin{equation}
%\hat\varphi \equiv \frac{\hat\Phi}{\Phi_0}=\phi(\hat a+\hat a^\dagger),\qquad 
%\Phi_0=\frac{h}{e},
%\end{equation}
% where $\phi$ is the LC flux enclosed by the QR.  
 
 %At energies below ~$\sim 1$ eV, graphene is described near the two valleys $\tau=\pm$ by the
%massless Dirac Hamiltonian. \textcolor{red}{Where is this Hamiltonian? Where is $\tau$ and $\sigma$ and $s$?}

%where $\boldsymbol{\sigma}$ acts in the A/B sublattice (pseudospin) space, $s$ is real spin,and 
We consider a narrow graphene ring of radius $R$, coupled to a quantized flux as above. The flux is added to the kinetic momentum via minimal coupling
\begin{equation}
\hat{\boldsymbol{\Pi}}=-i\hbar\nabla+e\,\mathbf A(\hat\Phi), \, 
A_\theta(r)=\frac{\hat\Phi}{2\pi r}, \, 
\Pi_\theta=-\frac{i\hbar}{r}\partial_\theta+\frac{e\hat\Phi}{2\pi r},
\nonumber 
\end{equation}
where $\theta$ is an angular coordinate on the ring. Projecting to the lowest radial mode of a narrow ring ($r\simeq R$) yields an effective
1D Dirac ring Hamiltonian
\begin{equation}
\hat H_{\rm ring}(\hat\varphi)=
\frac{\hbar v_F}{R}\sum_{\tau,s}\int_0^{2\pi}\! d\theta\;
\hat\psi_{\tau s}^\dagger(\theta)\,
\Big[-i\,\sigma_\theta\,\partial_\theta+\sigma_\theta\,\hat\varphi\Big]\,
\hat\psi_{\tau s}(\theta) \nonumber 
\label{eq:H_Dirac_ring}
\end{equation}
with $\sigma_\theta\equiv -\sigma_x\sin\theta+\sigma_y\cos\theta$.
Diagonalizing again in the basis of angular-momentum states with quantum number $m$ gives Dirac subbands labeled by an integer $m$.
Near the Dirac points one obtains the 1D Dirac ring spectrum
(with the usual Berry-phase shift $\beta=\tfrac12$). Projecting to the conduction band, the electron Hamiltonian is
\begin{equation}
\hat H_e(\hat\varphi)=\sum_{m,\alpha}\varepsilon_0\big|m+\beta+\hat\varphi\big|\;c_{m\alpha}^\dagger \hat c_{m\alpha}, \, 
  \varepsilon_0\equiv \frac{\hbar v_F}{R}
\label{eq:H_e_proj}
\end{equation}
where $\alpha$ includes spin and valley indices. 

Near neutrality and for small flux fluctuations (no level crossings at the Dirac point for the occupied set),
\begin{equation}
\big|m+\beta+\hat\varphi\big|=\big|m+\beta\big|+\hat\varphi\,{\rm sgn}(m+\beta),
\label{eq:abs_expand}
\end{equation}
so that $\hat\varphi$ couples linearly to the chirality (persistent-current) operator
\begin{equation}
\hat{\mathcal J}\equiv \sum_{m,\alpha}{\rm sgn}(m+\beta)\,\hat n_{m\alpha}
=\sum_{m\ge 0,\alpha}\hat n_{m\alpha}-\sum_{m\le -1,\alpha}\hat n_{m\alpha}, \nonumber 
\label{eq:J_def}
\end{equation}
where 
$\hat n_{m\alpha}=c_{m\alpha}^\dagger \hat c_{m\alpha}$. 
Including the LC mode gives
\begin{equation}
\hat H=\sum_{m,\alpha}\varepsilon_0|m+\beta|\,\hat n_{m\alpha}+\hbar\omega\Big(\hat a^\dagger\hat a+\tfrac12\Big)
+\varepsilon_0\phi\hat{X} \hat{\mathcal J}. 
\label{eq:H_linear}
\end{equation}

\paragraph{Mode elimination and effective interaction.}
In the strict low-energy Dirac theory (no explicit $A^2$ term), the oscillator can be eliminated exactly
by a displacement $\hat a\to \hat b-\lambda  \hat{\mathcal J}/(\hbar\omega)$, $\lambda=\varepsilon_0\phi$, yielding
\begin{equation}
\hat H= \sum_{m,\alpha}\varepsilon_0|m+\beta|\,\hat n_{m\alpha}+\hbar\omega\Big(\hat b^\dagger\hat b+\tfrac12\Big)-\chi_D(\phi)\,\hat{\mathcal J}^2 \nonumber 
\end{equation}
where $\chi_D(\phi)=\frac{\lambda^2}{\hbar\omega}.$ In a lattice-regularized (gauge-invariant) treatment, an
additional diamagnetic stiffness term $\propto (\hat a+\hat a^\dagger)^2 \hat D$ should be included. As shown in the SM, the  expectation value 
$\langle \hat D\rangle = -\Big(\frac{\phi}{N_s}\Big)^2 \langle \hat H_0\rangle$, where $N_s$ is the number of lattice sites on the ring.  Approximating $\hat D$ by its expectation value $D_{\rm eff}\equiv \langle \hat D\rangle$ in the projected low-energy sector gives a renormalized,
saturating coupling,  
\begin{equation}
\chi_{\rm lat}(\phi)=\frac{\lambda^2}{\hbar\omega+2D_{\rm eff}\phi^2}. 
\label{eq:chi_lat}
\end{equation}

%\qquad
%\chi_{\rm lat}(\phi\to\infty)\to %\frac{\varepsilon_0^{\,2}}{2D_{\rm %eff}}.

%This system doesn't assume the two-level approximation of matter and the coupling is through magnetic (transverse) field which is usually ignored in the typical treatment of coupling matter to light in cavity QED setups. 

\paragraph{Balanced versus polarized ground state and the critical point.}
We fix the (large) number of conduction electrons to $N$ and define $N_\pm$ as the occupations of the $m\ge 0$ and $m\le -1$
branches, respectively, so that $N=N_++N_-$ and $\mathcal J\equiv N_+-N_-$.
For large $N$, the minimal kinetic energy at given $(N_+,N_-)$ is obtained by filling consecutive modes on each branch,
each with capacity $g_d$; using $\sum_{r=1}^{q}(r-\tfrac12)=q^2/2$ gives $E_{\rm kin}(\mathcal J)\simeq \frac{\varepsilon_0}{4g_d}\Big(N^2+\mathcal J^2\Big),$
where $g_d$ is spin-valley degeneracy, equal to 4 for monolayer graphene. 
The effective electronic ground-state functional is therefore
\begin{equation}
E_{\rm eff}(\mathcal J)\simeq \frac{\varepsilon_0}{4g_d}N^2+\Big(\frac{\varepsilon_0}{4g_d}-\chi\Big)\mathcal J^2,
\label{eq:Eeff_J}
\end{equation}
with $\chi=\chi_D(\phi)$ or $\chi_{\rm lat}(\phi)$.
For $\chi<\chi_c$ the minimum is at the balanced configuration $\mathcal J\simeq 0$, while for $\chi>\chi_c$
the energy is minimized by the largest allowed $|\mathcal J|$ (cutoff-limited polarized state), implying a first-order
jump once the bound on $\mathcal J$ is enforced. The critical coupling is
\begin{equation}
\chi_c=\frac{\varepsilon_0}{4g_d } 
\; \Rightarrow\;
\frac{\hbar v_F}{16R}\ \ ({\rm at}\; g_d = 4).
\label{eq:chi_c}
\nonumber
\end{equation}
In the strict Dirac case, the corresponding critical flux amplitude is
\begin{equation}
\phi_c^2=\frac{\hbar\omega}{4g_d\,\varepsilon_0} 
\; \Rightarrow\; 
\frac{\omega R}{16 v_F}\ \ (g_d=4).
\label{eq:phi_c_dirac}
\end{equation}
For the lattice-regularized case \eqref{eq:chi_lat}, the transition exists provided $\chi_{\rm max}>\chi_c$
(i.e., $4g_d\,\varepsilon_0>2D_{\rm eff}$), and the critical flux amplitude is 
\begin{equation}
\phi_c^2=\frac{\hbar\omega}{4g_d\,\varepsilon_0-2D_{\rm eff}}.
\label{eq:phi_c_lat}
\end{equation}
Here  $\varepsilon_0$ and $\omega$ are tunable parameters, whereas $D_{\rm eff}$ is typically an order of magnitude smaller than $\varepsilon_0$, so including it will shift the critical value but won't prevent the transition. 

\paragraph{ Flux displacement across the transition.}
As described in the Appendix, for electrons on a lattice the diamagnetic term $\tfrac{\phi^2}{2}(\hat a+\hat a^\dagger)^2 \hat D$ needs to be added to the coupled Dirac-ring Hamiltonian (\ref{eq:H_linear}) in the projected conduction band. 
This implies that the LC mode is \emph{displaced} whenever the electronic ground state develops a nonzero
chirality imbalance $\langle\hat{\mathcal J}\rangle\neq0$. Using
$\hbar\omega(\hat a^\dagger\hat a+\tfrac12)=\frac{\hbar\omega}{4}(\hat X^2+\hat P^2)$ with
$\hat P=i(\hat a^\dagger-\hat a)$, the oscillator potential in a fixed electronic sector is
\begin{equation}
V(\hat X)=\Big(\frac{\hbar\omega}{4}+\frac{\phi^2}{2}D_{\rm eff}\Big)\hat X^2+\lambda \hat X\,\langle\hat{\mathcal J}\rangle,
\label{eq:Vx}
\end{equation}
where  $D_{\rm eff}=0$ in the strict low-energy Dirac theory. 
Minimization gives the ground-state displacement
\begin{equation}
{
\langle \hat a\rangle=\frac{\langle \hat X\rangle}{2}
=-\frac{\lambda}{\hbar\omega+2\phi^2 D_{\rm eff}}\;\langle\hat{\mathcal J}\rangle,
}
\label{eq:adispl}
\end{equation}
and the corresponding flux number in the displaced vacuum,
\begin{equation}
{
\langle \hat a^\dagger \hat a\rangle
=\big|\langle \hat a\rangle\big|^2
=\frac{\lambda^2}{(\hbar\omega+2\phi^2 D_{\rm eff})^2}\;\langle\hat{\mathcal J}\rangle^2.
}
\label{eq:nph}
\end{equation}

In the balanced phase, $\langle\hat{\mathcal J}\rangle\simeq 0$, hence
$\langle \hat a\rangle\simeq 0$ and the LC field has no macroscopic displacement.
For $\chi>\chi_c$ the minimum of the electronic free energy moves to the cutoff-limited polarized sector,
$|\langle\hat{\mathcal J}\rangle|=\mathcal J_{\max}$, implying a \emph{discontinuous jump}, 
\begin{equation}
\langle \hat a\rangle = 0\ \Rightarrow\
\mp\,\frac{\lambda\,\mathcal J_{\max}}{\hbar\omega+2\phi^2 D_{\rm eff}}, \,
\langle \hat a^\dagger\hat a\rangle \Rightarrow\
\frac{\lambda^2\mathcal J_{\max}^2}{(\hbar\omega+2\phi^2 D_{\rm eff})^2},
\nonumber
\end{equation}
consistent with a first-order (level-crossing) transition.
The Hamiltonian is invariant under the $Z_2$ transformation
$\hat a\!\to\!-\hat a$ and $\hat{\mathcal J}\!\to\!-\hat{\mathcal J}$, so the polarized phase has two degenerate
branches $\pm\mathcal J_{\max}$ with opposite values of $\langle\hat a\rangle$. 
%; in practice an infinitesimal bias flux selects one branch.

%%%%%%%%%%%%%%%%%%%%%%%%%%%%%%%%

\section{The nonlinear cavity}

Beyond the linear resonator, introducing JJ to the plain LC circuit will result in a Kerr nonlinearity $\alpha_4(\hat a+\hat a^\dagger)^4$ which can be incorporated nonperturbatively. This provides an analytically anchored pathway to explore nonlinear flux-matter phases and sector-selective photon dressing in regimes relevant to circuit QED and mesoscopic rings~\cite{Blais2021,Manucharyan2009,Mooij1999}, and complementary to cavity-mediated pairing proposals~\cite{Schlawin2019}. The details are in the Appendix. One interesting consequence of the nonlinearity is that the resulting dressed mode frequency $\Omega$ depends nontrivially on the electron configuration, i.e., electron distribution over angular momentum states. Therefore, one can control it by changing the electron distribution.

Here we would like to focus on an interesting possibility of a {\it synthetic nonlinearity}. 
 
%%%%%%%%%%%%%%%%%%%%%%%%%%%%%

\subsection{Synthetic Josephson junction in tight-binding systems} 

Here we show that interactions induced in a tight binding ring by coupling to a linear LC circuit give rise to the nonlinearity equivalent to the synthetic JJ. By exactly diagonalizing the model we obtain that the resulting Hamiltonian for the LC circuit is isomorphic to the Hamiltonian of a circuit with a JJ. Since the synthetic JJ element emerges due to electron interactions in a ring,  the JJ parameters that are normally fixed by fabrication can be tuned by controlling the QR degrees of freedom.

%%%%%%%%%%%%%%%%%%%%%%%%%%%%%%%%%

\subsubsection{Model and sector reduction at arbitrary coupling}

For a lattice with creation/annihilation operators $\hat c_i^\dagger,\hat c_i$,
nearest-neighbor hopping amplitudes $t_{ij}$, and coupling to the vector potential $\mathbf A(\mathbf r)$,
the Peierls substitution prescribes
\begin{equation}
t_{ij} \rightarrow t_{ij}\,e^{\,i\theta_{ij}} = t_{ij}\,
\exp\!\left(\frac{i q}{\hbar}\int_{\mathbf r_i}^{\mathbf r_j}\mathbf A\cdot d\boldsymbol\ell\right),
\label{eq:peierls}
\end{equation}
so the tight-binding Hamiltonian for the electrons becomes
\begin{equation}
\hat H \;=\; -\sum_{\langle i,j\rangle}
\Big[t_{ij}\,e^{\,i\theta_{ij}}\;\hat c_i^\dagger \hat c_j + \text{H.c.}\Big]. 
\label{eq:H_tb}
\end{equation}

%with the bond phase
%\begin{equation}
%\theta_{ij}\;=\;\frac{q}{\hbar}\int_{\mathbf r_j}^{\mathbf r_i}\mathbf A\cdot d\boldsymbol\ell .
%\label{eq:theta_ij}
%\nonumber 
%\end{equation}

On a ring where each bond shares the same line integral, we define
\begin{equation}
\eta \;\equiv\; \frac{q}{\hbar}\,\frac{\Phi}{M}
\;=\; \frac{2\pi}{M}\,\frac{\Phi}{\Phi_0},
\; 
\Phi=\oint \mathbf A\cdot d\boldsymbol\ell,\; 
\Phi_0=\frac{h}{q}.
%\label{eq:eta_def}
\nonumber 
\end{equation}
Then Eq.~\eqref{eq:H_tb} reduces to
\begin{equation}
\hat H \;=\; -t\sum_{j=1}^{M}
\Big(e^{\,i\eta}\,\hat c_{j+1}^\dagger \hat c_j + e^{-\,i\eta}\,\hat c_{j}^\dagger \hat c_{j+1}\Big),
\label{eq:H_ring}
\end{equation}
and for a momentum with periodic boundary conditions ($k=\tfrac{2\pi}{Ma}n$) one has 
\begin{equation}
\hat H \;=\; \sum_{k}\, \varepsilon(k-\eta/a)\;\hat c_k^\dagger \hat c_k,
\quad
\varepsilon(q)=-2t\cos(qa),
\label{eq:H_k}
\end{equation}
where $a$ is the lattice period.

Consider the Peierls phase factor when the phase is quantized due to the quantized LC resonator field as above,  $e^{i\eta\hat X}$ on each bond, with the previously introduced bosonic quadratures. Substituting
$$\hat c_k=\frac{1}{\sqrt M}\sum_j e^{-ikja}\hat c_j $$ 
with $k=\tfrac{2\pi}{Ma}n$ ($n=0,\dots,M-1$) yields the block Hamiltonian in each occupation sector $\{n_k\}$:
\begin{equation}
\hat H_{\{n_k\}}=
\sum_k\!\big[-2t\cos(ka-\eta\hat X)\big]\hat n_k
+\hbar\omega\,\hat a^\dagger\hat a
,
\label{eq:Hsector}
\end{equation}
with $\hat n_k=\hat c_k^\dagger\hat c_k$ \emph{commuting} with $\hat a,\hat a^\dagger$. Define
\begin{equation}
C=\sum_k n_k\cos(ka),\; 
S=\sum_k n_k\sin(ka),
\end{equation}
so that Eq.~\eqref{eq:Hsector} is exactly
\begin{equation}
\hat H_{\{n_k\}}
= -2t\!\left[C\cos(\eta\hat X)-S\sin(\eta\hat X)\right]
+\hbar\omega\,\hat a^\dagger\hat a.
\label{eq:HCS}
\end{equation}
Here the sums over the number of particles give constants.
No approximation has been made: the many-body problem factorizes into disjoint fermionic sectors and a \emph{single} bosonic mode. The procedure for its numerical diagonalization is outlined in the Appendix. 

%%%%%%%%%%%%%%%%%%%%%%%%%%%%%%%%%

\subsubsection{Flux-qubit (rf-SQUID) mapping}

Compressing 
$ C\cos(\eta\hat X)-S\sin(\eta\hat X)
=\sqrt{C^2+S^2}\,\cos(\eta\hat X+\delta)$,
where $\tan\delta= S/C$, we introduce the superconducting phase and its conjugate ``charge'',
\begin{equation}
\hat\phi\equiv \eta \hat X+\delta,\quad
\hat n_\phi\equiv \frac{\hat P}{\eta},\quad
[\hat\phi,\hat n_\phi]=i,
\end{equation}
and note $\hbar\omega\,\hat a^\dagger\hat a=\frac{\hbar\omega}{2}(\hat P^2+\hat X^2)+\text{const}$. Up to an additive constant, each sector $\{n_k\}$ maps \emph{exactly} to the rf-SQUID Hamiltonian
\begin{equation}
{\;
\hat H_{\{n_k\}}
= 4E_C\,\hat n_\phi^2+\frac{E_L}{2}\,(\hat\phi-\phi_{\rm ext})^2
- E_J\cos\hat\phi,
\;}
\label{eq:rfSQUID}
\end{equation}
with the parameter identification
\begin{equation}
E_J=2t\,\sqrt{C^2+S^2},\; 
\phi_{\rm ext}=\delta, \; 
E_L=\frac{\hbar\omega}{\eta^2},\; 
E_C=\frac{\hbar\omega\,\eta^2}{8}.
\nonumber 
%\label{eq:map2}
\end{equation}
Two immediate consequences are: (i) the geometric mean is fixed,
$
E_C E_L=\displaystyle \frac{(\hbar\omega)^2}{8},
$
independent of $\eta$, while $\eta$ trades charge-like vs flux-like regimes; and (ii) the standard rf-SQUID double-well criterion is governed by
$$
\beta\equiv \frac{E_J}{E_L} \displaystyle 
=\frac{2t\,\eta^{2}}{\hbar\omega}\,\sqrt{C^2+S^2},
$$ 
so near the bias value $\phi_{\rm ext}\approx\pi$ a double-well forms for $\beta\gtrsim 1$, yielding the familiar two-level flux-qubit manifold. Tuning $C$ and $S$ can control the $\beta$ parameter. \\

\section{Conclusions}

Solid-state platforms that coherently couple electrons to the quantized flux of superconducting circuits provide a versatile setting to explore emergent collective behavior, where circuit-mediated interactions intertwine with intrinsic electronic correlations in qualitatively new ways.

We introduced a gauge-invariant, analytically tractable and tunable platform based on cavity quantum magnetostatics, where quantized magnetic flux from an LC resonator couples minimally to the orbital motion of fermions. This coupling generates a collective, attractive current-current interaction between fermions, which is strong enough to realize a rich variety of collective phenomena, such as an orbital Stoner-Dicke instability or quantum phase transition, while crucially retaining the diamagnetic term and avoiding the RWA. As a result, the model supports a first-order fermionic superradiant transition without sacrificing gauge invariance, highlighting a qualitatively distinct route to this type of quantum phase transitions compared to electric-dipole cavity QED systems. Near criticality, the system  can be used as a magnetometer since the state of the field is squeezed. 

Adding the JJ nonlinearity opens access to genuinely nonlinear flux–matter phases, including sector-selective photon dressing, in experimentally relevant circuit-QED regimes. Finally, our tight-binding realization of a synthetic, tunable JJ nonlinearity shows that key aspects of the nonlinear physics can be engineered without physical junctions, broadening the set of viable experimental implementations. These results establish a compact framework for exploring controllable, flux-mediated many-body physics and motivate future work on nonequilibrium dynamics, finite-geometry effects, and ordered phases induced by orbital current–current interactions. 

%%%%%%%%%%%%%%%%%%%%%%%%%%%%%%%%%%%
{\bf Acknowledgments.} This work has been supported in part by the Keck Foundation (Award No.~CRM:0132347) and by the Laboratory Directed Research and Development program and Sandia University Partnerships Network ( SUPN) program and performed in part at the Center for Integrated Nanotechnologies ( CINT), an Office of Science User Facility operated for the U.S. Department of Energy (DOE) Office of Science. Sandia National Laboratories is a multimission laboratory managed and operated by National Technology and Engineering Solutions of Sandia, LLC., a wholly owned subsidiary of Honeywell International, Inc., for the U.S. Department of Energy's National Nuclear Security Administration under Contract No. DE-NA0003525. This article describes objective technical results and analysis. Any subjective views or opinions that might be expressed in the article do not necessarily represent the views of the U.S. Department of Energy or the United States Government.

%%%%%%%%%%%%%%%%%%%%%%%%%%%%%%%%%%%%%%%%%%%%

\appendix

\section{Fermionic Phase Transition}

%We now extend the analysis to a system of $N$ spinless fermions. Due to the Pauli exclusion principle, fermions cannot condense into a single momentum state. Instead, the ground state is a Fermi sea. Surprisingly, we demonstrate that the condition for the onset of the superradiant current remains identical to the bosonic case due to the decoupling of the center-of-mass motion in a parabolic band.

\subsection{The Balanced Phase}
In the normal balanced phase ($M=0$), the $N$ fermions occupy the lowest available kinetic energy states symmetrically around zero angular momentum. Assuming $N$ is odd, the occupied states range from $m = -k_F$ to $m = +k_F$, where $k_F = (N-1)/2$.

The total angular momentum vanishes by symmetry:
\begin{equation}
M = \sum_{i=1}^N m_i = 0.
\end{equation}
The total kinetic energy of this ground state is:
\begin{equation}
E_{\text{bal}} = g_{\text{eff}} \sum_{i=1}^N m_i^2.
\end{equation}
This term represents the internal energy of the Fermi sea.

\subsection{The Polarized Phase }
The ``polarized'' state for fermions corresponds to a rigid shift of the entire Fermi distribution in momentum space. We consider a variational state where every particle is boosted by one unit of angular momentum: $m_i \to m_i + 1$.

The new total angular momentum is 
\begin{equation}
M_{\text{pol}} = \sum_{i=1}^N (m_i + 1) = \sum_{i=1}^N m_i + \sum_{i=1}^N 1 = 0 + N = N.
\end{equation}
This yields the same macroscopic momentum $M=N$ as in the bosonic case.

The kinetic energy of the shifted state is
\begin{equation}
E'_{\text{kin}} = g_{\text{eff}} \sum_{i=1}^N (m_i + 1)^2 = g_{\text{eff}} \left( \sum m_i^2 + 2\sum m_i + \sum 1 \right).
\nonumber 
\end{equation}
Using $\sum m_i = 0$ and $\sum 1 = N$, this simplifies to
\begin{equation}
E'_{\text{kin}} = E_{\text{bal}} + g_{\text{eff}} N.
\nonumber 
\end{equation}
The interaction energy, which depends only on the total angular momentum $M=N$, is
\begin{equation}
E_{\text{int}} = - \frac{(2g\phi N)^2}{\hbar\omega + 4gN\phi^2}.
\nonumber 
\end{equation}
The total energy of the polarized phase is therefore 
\begin{equation}
E_{\text{pol}} = E_{\text{bal}} + g_{\text{eff}} N - \frac{4g^2\phi^2 N^2}{\hbar\omega + 4gN\phi^2}.
\end{equation}

\subsection{The Phase Transition Condition}
The system undergoes a transition when the energy of the polarized phase becomes lower than that of the balanced phase, $E_{\text{pol}} < E_{\text{bal}}$. Substituting the expressions above, 
\begin{equation}
E_{\text{bal}} + g_{\text{eff}} N - \frac{4g^2\phi^2 N^2}{\hbar\omega + 4gN\phi^2} < E_{\text{bal}}.
\end{equation}
Crucially, the internal Fermi energy $E_{\text{bal}}$ cancels out exactly. The inequality simplifies to
\begin{equation}
g_{\text{eff}} N - \frac{4g^2\phi^2 N^2}{\hbar\omega + 4gN\phi^2} < 0.
\nonumber 
\end{equation}
Dividing by $N$, we obtain the critical condition:
\begin{equation}
g_{\text{eff}} < \frac{4g^2\phi^2 N}{\hbar\omega + 4gN\phi^2}.
\label{eq:critical_condition}
\end{equation}
We can invert Eq.~(\ref{eq:critical_condition}) to find the critical mode function parameter $\phi_c$ required to trigger the transition. Solving for $\phi$ at the phase boundary yields 
\begin{equation}
\phi_c = \sqrt{\frac{g_{\text{eff}}\hbar\omega}{4gN (g - g_{\text{eff}})}}.
\label{eq:phi_c}
\end{equation}
This solution implies that a phase transition is only possible if the bare cavity coupling $g$ exceeds the effective lattice stiffness $g_{\text{eff}}$.

The derived condition is identical to the bosonic result. Physically, even though fermions possess a large internal kinetic energy (Fermi pressure), the infinite-range cavity interaction couples only to the total momentum operator $M$. In a system with parabolic dispersion (harmonic trap or ring), the center-of-mass mode is mathematically independent of the relative motion of the particles (Kohn's theorem). Consequently, the ``stiffness'' of the system against a global boost, or the energy cost to accelerate the entire cloud from $M=0$ to $M=N$, is simply $N g_{\text{eff}}$, independent of the quantum statistics.

\section{Exact diagonalization of the model}

Here we show that the model represented by the  Hamiltonian (5) in the main text is exactly diagonalizable by a conditional displacement followed by squeezing Bogoliubov transformations, yielding a renormalized normal mode of frequency $\Omega$ and an induced all-to-all attractive interaction $-\chi\,\hat M^2$ in the matter sector, where the conserved total angular momentum
\begin{equation}
\hat M=\sum_{i=1}^N \hat L_i,\qquad
[\hat M,\hat a]=[\hat M,\hat a^\dagger]=0.
\label{eq:Ltot}
\end{equation}

We start by introducing dimensionless quadratures
$\hat x=(\hat a+\hat a^\dagger)/\sqrt2$, $\hat p=(\hat a^\dagger-\hat a)/(i\sqrt2)$,
so that $\hat a^\dagger\hat a=\tfrac12(\hat x^2+\hat p^2-1)$ and
$(\hat a^\dagger+\hat a)^2=2\hat x^2$. Substituting into Eq.~(5) from the main text yields 
\begin{align}
\hat H_0 = g_{\rm eff}\sum_{i=1}^N \hat L_i^2
+\frac{\hbar\omega}{2}\hat p^2
+ A\,\hat x^2
+ B\,\hat x
-\frac{\hbar\omega}{2},
\label{eq:Hquad}
\end{align}
with
\begin{equation}
A=2g\phi^2 N+\frac{\hbar\omega}{2},\qquad
B=-2\sqrt2\,g\phi\,\hat M.
\label{eq:AB}
\end{equation}

\subsection{Displacement}

Next, we complete the square: $A\hat x^2+B\hat x=A(\hat x-\hat x_0)^2-\tfrac{B^2}{4A}$ with
\begin{equation}
\hat x_0=\frac{-B}{2A}=\frac{\sqrt2\,g\phi}{A}\,\hat M,\qquad
\frac{B^2}{4A}=\frac{2g^2\phi^2}{A}\,\hat M^2.
\label{eq:x0}
\end{equation}
Since $[\hat p,\hat M]=0$, the unitary operation $\hat U_{\rm disp}=e^{-i\hat p\,\hat x_0}$ removes the linear term:
\begin{equation}
\hat U_{\rm disp}^\dagger \hat H_0 \hat U_{\rm disp}
= g_{\rm eff}\sum_{i=1}^N \hat L_i^2
+\frac{\hbar\omega}{2}\hat p^2
+ A\,\hat x^2
-\frac{2g^2\phi^2}{A}\,\hat M^2
-\frac{\hbar\omega}{2}.
\nonumber 
%\label{eq:Hdisp}
\end{equation}

\subsection{Squeezing} 

The quadratic oscillator part has the form
$H_f=\tfrac{\alpha}{2}\hat p^2+\tfrac{\beta}{2}\hat x^2$ with
\begin{equation}
\alpha=\hbar\omega,\qquad
\beta=2A=4g\phi^2 N+\hbar\omega.
\label{eq:alphabeta}
\end{equation}
Substituting 
\begin{align}
\hat x^2&=\tfrac12\!\left(\hat a^2+\hat a^{\dagger 2}\right)+\hat a^\dagger\hat a+\tfrac12,\nonumber  \\
\hat p^2&=-\tfrac12\!\left(\hat a^2+\hat a^{\dagger 2}\right)+\hat a^\dagger\hat a+\tfrac12, \nonumber 
\end{align}
yields the quadratic bosonic form
\begin{align}
\hat H_f
&=\frac{\alpha+\beta}{2}\,\hat a^\dagger\hat a
+\frac{\beta-\alpha}{4}\!\left(\hat a^2+\hat a^{\dagger 2}\right)
+\frac{\alpha+\beta}{4}\nonumber
\end{align}

After applying the Bogoliubov (squeezing) transformation,
\begin{equation}
\hat a=\cosh r\,\hat b+\sinh r\,\hat b^\dagger,\text{ where } 
\tanh(2r)=\frac{\beta-\alpha}{\beta+\alpha},
\nonumber 
\end{equation}
or, equivalently, 
\begin{equation}
r=\frac14\ln\!\frac{\beta}{\alpha},
\nonumber 
\end{equation}
we obtain the noninteracting Hamiltonian: 
\begin{equation}
\hat H_f=\varepsilon\left(\hat b^\dagger\hat b+\tfrac12\right),\qquad
\varepsilon=\sqrt{\alpha\,\beta}.
\nonumber 
\end{equation}

This can be also achieved by rescaling of quadratures,
\begin{equation}
\tilde x=\Bigl(\frac{\beta}{\alpha}\Bigr)^{\!1/4}\hat x,\qquad
\tilde p=\Bigl(\frac{\alpha}{\beta}\Bigr)^{\!1/4}\hat p,
\nonumber 
\end{equation}
which puts the Hamiltonian into the isotropic form
\begin{equation}
\hat H_f=\frac12\sqrt{\alpha\beta}\,\bigl(\tilde p^{\,2}+\tilde x^{\,2}\bigr).
\nonumber 
\end{equation}

The energy spacing  (or the oscillator frequency) reads
\begin{equation}
\varepsilon = \hbar \Omega = \sqrt{\hbar\omega\bigl(4g\phi^2N+\hbar\omega\bigr)}. 
\nonumber 
\end{equation}

The diagonalizing unitary transformation is equivalent to the single-mode squeeze operator
\begin{equation}
S(r)=\exp\!\Bigl[\frac{r}{2}\bigl(\hat a^{2}-\hat a^{\dagger 2}\bigr)\Bigr],\quad
\hat b=S^\dagger(r)\,\hat a\,S(r),
\nonumber 
\end{equation}
which acts on the canonically conjugate quadratures as
\begin{equation}
S^\dagger(r)\,\hat x\,S(r)=e^{-r}\hat x,\quad
S^\dagger(r)\,\hat p\,S(r)=e^{+r}\hat p.
\nonumber 
\end{equation}
Thus, in the $\hat b$-vacuum,
\begin{equation}
\langle \hat x^2\rangle=\tfrac12 e^{-2r},\quad
\langle \hat p^2\rangle=\tfrac12 e^{+2r},\quad
\langle \hat x^2\rangle\,\langle \hat p^2\rangle=\tfrac14,
\nonumber 
\end{equation}
showing that one quadrature variance is reduced below the vacuum level while the conjugate one is increased by the same amount, achieving an ideal squeezing. The anisotropy in $\hat H_f$ (i.e., $\beta\neq\alpha$) precisely demands this squeeze; indeed,  $r=\tfrac14\ln(\beta/\alpha)$.

Equivalently, the rescaling above makes the dynamics isotropic, with
\begin{equation}
\langle \hat x^2\rangle=\tfrac12\sqrt{\frac{\alpha}{\beta}},\quad
\langle \hat p^2\rangle=\tfrac12\sqrt{\frac{\beta}{\alpha}}.
\nonumber 
\end{equation}
Since $\beta>\alpha$, the position is squeezed and the momentum is anti-squeezed. Physically, the $\hat a^2+\hat a^{\dagger 2}$ term, which  originates from the diamagnetic term, is the degenerate parametric (two-photon) interaction used to generate squeezed states \cite{Scully1997}.

The total Hamiltonian reads 
\begin{equation}
\hat S^\dagger \hat U_{\rm disp}^\dagger \hat H_0 \hat U_{\rm disp}S
= g_{\rm eff}\sum_{i=1}^N \hat L_i^2
-\chi\,\hat M^2
+\hbar\Omega\Big(\hat b^\dagger\hat b+\tfrac12\Big)
-\frac{\hbar\omega}{2},
\label{eq:Hdiag}
\end{equation}
with the induced collective coupling
\begin{equation}
\chi=\frac{2g^{2}\phi^{2}}{A}
=\frac{4g^{2}\phi^{2}}{\,4g\phi^{2}N+\hbar\omega\,}.
\label{eq:chi}
\end{equation}

Because all $\hat L_i$ commute, we can choose simultaneous eigenstates
$\ket{\{m_i\}}$ with $m_i\in\mathbb{Z}$ and $M=\sum_i m_i$.
Eigenenergies factorize as
\begin{equation}
E(\{m_i\},n)
= g_{\rm eff}\sum_{i=1}^N m_i^2
-\chi\,M^2
+\hbar\Omega\Big(n+\tfrac12\Big)
-\frac{\hbar\omega}{2},
\label{eq:Efull}
\end{equation}
where $n=0,1,\dots$ and the corresponding eigenstates are displaced--squeezed harmonic oscillator states tensored with $\ket{\{m_i\}}$.
For the ground state one may set $n=0$, since the last two terms in \eqref{eq:Efull} are configuration independent.
The squeezing level $r$ can be controlled by changing the $\phi$-flux enclosed by the ring, or by changing $N$, for example by gating.
%%%%%%%%%%%%%%%%%%%%%%%%%%%%%%%%%%%%%

\section{Zeeman coupling to electron spins}

Here we include the coupling of the spin degree of freedom of electrons
moving on a ring to the same quantized magnetic flux that enters the orbital
minimal coupling. Writing the cavity quadrature as
$\hat X\equiv \hat a+\hat a^\dagger$,
the Hamiltonian reads
\begin{equation}
\hat H= g_{\rm eff}\sum_{i=1}^N \hat L_i^{\,2}
+\hbar\omega\,\hat a^\dagger\hat a
+ gN\phi^2 \hat X^2
-\hat X\Big(2g\phi\,\hat M+\eta\,\hat\Sigma_z\Big)
\label{eq:Hspin}
\end{equation}
 where 
\begin{equation}
\hat M \equiv \sum_{i=1}^N \hat L_i,
\qquad
\hat\Sigma_z \equiv \sum_{i=1}^N \sigma_i^z,
\label{eq:conserved}
\end{equation}
with the compact Zeeman coupling
\begin{equation}
\eta \equiv \frac{g_s\mu_B}{2}B_{\rm zpf}.
\label{eq:eta}
\end{equation}
where $g_s$ is the electron $g$-factor, $\mu_B$ the Bohr magneton,($B_{\rm zpf}$)
denote the static (zero-point) magnetic-field components normal to the ring. Since $[\hat H,\hat M]=[\hat H,\hat\Sigma_z]=0$, the diagonalization proceeds independently in each
sector of fixed $(M,\Sigma_z)$.

The cavity mode experiences an $\hat X^2$-stiffening and an $(M,\Sigma_z)$-dependent linear drive.
After a Bogoliubov (squeezing) transformation and a subsequent displacement, the Hamiltonian becomes
\begin{equation}
\hat H
= g_{\rm eff}\sum_{i=1}^N \hat L_i^{\,2}
-\frac{\hat J^{\,2}}{\hbar\omega+4gN\phi^2}
+\hbar\Omega\Big(\hat b^\dagger\hat b+\tfrac12\Big)
\;+\;{\rm const.},
\label{eq:Hdiag}
\end{equation}
where the dressed cavity frequency is
\begin{equation}
\Omega=\omega\sqrt{1+\frac{4gN\phi^2}{\hbar\omega}},
\label{eq:Omega}
\end{equation}
and the conserved operator driving the cavity is
\begin{equation}
\hat J \equiv 2g\phi\,\hat M+\eta\,\hat\Sigma_z.
\label{eq:Jdef}
\end{equation}
Equation~\eqref{eq:Hdiag} exhibits an exact cavity-mediated interaction
$-\hat J^{\,2}/(\hbar\omega+4gN\phi^2)$, which contains orbital--orbital, spin--spin, and mixed
spin--orbital interaction terms upon expanding $\hat J^2$.

For an electronic configuration $\ket{\{m_i,s_i\}}$ with
$\hat M\ket{\{m_i,s_i\}}=M\ket{\{m_i,s_i\}}$ and
$\hat\Sigma_z\ket{\{m_i,s_i\}}=\Sigma\ket{\{m_i,s_i\}}$
($M=\sum_i m_i$, $\Sigma=N_\uparrow-N_\downarrow$),
the exact spectrum factorizes into the dressed cavity ladder and a matter-dependent shift,
\begin{equation}
E_{n,\{m_i,s_i\}}
=
g_{\rm eff}\sum_{i=1}^N m_i^2
-\frac{\big(2g\phi\,M+\eta\,\Sigma\big)^2}{\hbar\omega+4gN\phi^2}
+\hbar\Omega\Big(n+\tfrac12\Big),
\label{eq:spectrum}
\end{equation}
with $n=0,1,2,\dots$ labeling excitations of the displaced-squeezed cavity mode $\hat b$.

\subsection{Spin and orbital coupling interplay}

We analyze the stability of the balanced ground state defined by $\langle M \rangle = 0$ and $\langle \Sigma \rangle = 0$. We introduce a composite order parameter vector $\boldsymbol{\psi} = (M, \Sigma)^T$. The effective potential energy surface $V(\boldsymbol{\psi})$ near the origin is derived by expanding the energy functional $E_{n,\{m_i,s_i\}}$ to quadratic order in the order parameters.

The total energy consists of the single-particle confinement cost and the collective cavity-mediated interaction:
\begin{equation}
    E(\boldsymbol{\psi}) = E_{\text{cost}}(\boldsymbol{\psi}) - \frac{\left( 2g\phi M + \eta \Sigma \right)^2}{\mathcal{D}},
    \label{eq:energy_functional}
\end{equation}
where $\mathcal{D} \equiv \hbar\omega + 4gN\phi^2$ is the renormalized cavity mode energy, which explicitly incorporates the collective diamagnetic shift.

The energy cost to form macroscopic orbital ($M$) and spin ($\Sigma$) polarization arises from the kinetic term $g_{\text{eff}}\sum m_i^2$. In the Fermi liquid regime, minimizing the kinetic energy $\sum m_i^2$ subject to fixed total $M$ and $\Sigma$ yields the stiffness coefficients $\chi_M^{-1} = 2g_{\text{eff}}/N$ and $\chi_\Sigma^{-1} = g_{\text{eff}}N/2$. The Hessian matrix of the system, $\mathcal{H}_{ij} = \partial^2 E / \partial \psi_i \partial \psi_j$, is therefore given by:
\begin{equation}
    \mathcal{H} = 
    \begin{pmatrix}
        \frac{2g_{\text{eff}}}{N} - \frac{8g^2\phi^2}{\mathcal{D}} & -\frac{4g\phi\eta}{\mathcal{D}} \\
        -\frac{4g\phi\eta}{\mathcal{D}} & \frac{g_{\text{eff}}N}{2} - \frac{2\eta^2}{\mathcal{D}}
    \end{pmatrix}.
\end{equation}
The system undergoes a phase transition when the lowest eigenvalue of $\mathcal{H}$ becomes negative. The phase boundary is defined by the condition $\det(\mathcal{H}) = 0$, which implies:
\begin{equation}
    \left( \frac{g_{\text{eff}}}{N} - \frac{4g^2\phi^2}{\mathcal{D}} \right) \left( \frac{g_{\text{eff}}N}{4} - \frac{\eta^2}{\mathcal{D}} \right) = \left( \frac{2g\phi\eta}{\mathcal{D}} \right)^2.
    \label{eq:det_zero}
\end{equation}
Multiplying Eq.~(\ref{eq:det_zero}) by $\mathcal{D}^2$ allows us to decouple the interaction terms. We observe that the cross-terms involving the product of orbital and spin couplings cancel exactly. Rearranging for the critical spin coupling $\eta_c$, we obtain:
\begin{equation}
    \eta_c^2 \left[ \frac{g_{\text{eff}}\mathcal{D}}{N} \right] = \left( \frac{g_{\text{eff}}\mathcal{D}}{N} - 4g^2\phi^2 \right) \frac{g_{\text{eff}}N\mathcal{D}}{4}, 
\end{equation}
or 
\begin{equation}
    \eta_c^2 = \frac{N^2}{4} \left[ g_{\text{eff}}(\hbar\omega + 4gN\phi^2) - 4g^2N\phi^2 \right].
\end{equation}
When the effective mass renormalization is negligible (i.e., $g_{\text{eff}} \approx g$), the diamagnetic contribution $4g^2N\phi^2$ exactly cancels the attractive orbital term inside the bracket. This cancellation is the manifestation of the textbook \textit{no-go theorem} for the orbital sector. However, due to the hybridization with the spin degree of freedom, the critical coupling reduces to a remarkably simple form:
\begin{equation}
    \eta_c = \frac{1}{2} \sqrt{g N \hbar\omega}.
    \label{eq:critical_coupling}
\end{equation}
This result highlights a crucial physical mechanism: the orbital degree of freedom $M$, effectively ``screens'' the diamagnetic shift for the spin sector. Consequently, the critical coupling $\eta_c$ scales with the bare cavity frequency $\omega$ rather than the diamagnetically renormalized frequency, significantly lowering the threshold for vacuum-induced ferromagnetism compared to a pure spin model.

Writing the critical value in terms of $\phi$ yields  
\begin{equation}
    \phi_c = \frac{1}{N} \sqrt{ \frac{\eta^2 - \frac{1}{4}N g_{\rm eff} \hbar \omega}{g(g_{\rm eff} - g)} }.
    \label{eq:critical_phi}
\end{equation}

When $\eta$ goes to zero one recovers Eq.!(\ref{eq:phi_c}). No finite Zeeman--like
coupling is required: the ground state is already unstable, driven by the orbital channel.

Crucially, at the onset of ordering (just above criticality), the emergent collective mode is
a locked spin--orbital combination rather than a purely spin or purely orbital deformation.
The corresponding soft eigenvector fixes the ratio of the induced total angular momentum
and spin polarization to be
\begin{equation}
\left.\frac{M}{\Sigma}\right|_{\eta=\eta_c}
=
\frac{(2g\phi)\,\eta_c}{(g_{\rm eff}D/N)-(2g\phi)^2},
\label{eq:cor2_locking_ratio}
\end{equation}
so that the ordered phase generically exhibits mixed spin--orbital character.

%%%%%%%%%%%%%%%%%%%%%%%%%%%%%%%%%%%%%

\section{Diamangetic term}

In a gauge-invariant tight-binding description of a 1D lattice ring, a spatially uniform quantum magnetic
flux enters through Peierls phases. For nearest-neighbor hopping $\varepsilon_0$ on an $N_s$-site ring,
\begin{equation}
\hat H_{\rm tb}(\hat\varphi)=
-\varepsilon_0\sum_{j=1}^{N_s}\Big(
e^{\,i\hat\varphi/N_s}\,\hat c_{j+1}^\dagger \hat c_j+\text{H.c.}\Big),
\qquad 
\hat\varphi=\phi\,\hat X,
\label{eq:Htb_ring}
\end{equation}
where $\hat\varphi$ is the dimensionless flux threading the ring and $\phi$ sets the quantum-flux amplitude.
Expanding in $\phi$ to second order yields
\begin{equation}
\hat H_{\rm tb}=\hat H_0(\varphi_{\rm ext})+\hat X\,\hat J(\varphi_{\rm ext})
+\frac{\hat X^2}{2}\,\hat D(\varphi_{\rm ext})+O(\phi^3),
\label{eq:expansion}
\end{equation}
with $\hat H_0(\varphi_{\rm ext})\equiv \hat H_{\rm tb}(\phi=0)$ and the diamagnetic (stiffness) operator
\begin{align}
\hat D(\varphi_{\rm ext})
&=\Big(\frac{\phi}{N_s}\Big)^2 \varepsilon_0\sum_{j=1}^{N_s}
\Big(e^{i\varphi_{\rm ext}/N_s}\hat c_{j+1}^\dagger \hat c_j+\text{H.c.}\Big)
\nonumber\\
&= -\Big(\frac{\phi}{N_s}\Big)^2 \hat H_0(\varphi_{\rm ext}),
\label{eq:D_operator}
\end{align}
where the last equality holds identically for the uniform-flux ring. Consequently, its expectation value is
locked to the kinetic energy,
\begin{equation}
\langle \hat D\rangle = -\Big(\frac{\phi}{N_s}\Big)^2 \langle \hat H_0\rangle .
\label{eq:D_expect}
\end{equation}
For noninteracting spinless fermions at filling $\nu=N/N_s$ (large $N_s$),
$\langle \hat H_0\rangle/N_s=-(2\varepsilon_0/\pi)\sin(\pi\nu)$, implying
\begin{equation}
\frac{\langle \hat D\rangle}{\varepsilon_0}
=\frac{2}{\pi}\,\frac{\phi^2}{N_s}\,\sin(\pi\nu),
\label{eq:D_over_eps0}
\end{equation}
(and twice this value for spin degeneracy). Thus, in the uniform-flux geometry the diamagnetic stiffness is
extensive but parametrically small in units of the bare hopping scale, scaling as
$\langle \hat D\rangle/\varepsilon_0\sim \phi^2/N_s$ at fixed filling.

%%%%%%%%%%%%%%%%%%%%%%%%%

%%%%%%%%%%%%%%%%%%%%%%%%%%%%%
\section{Nonlinear cavity}
We start from
\begin{equation}
\hat{H}_0
= g\sum_{i=1}^{N}\!\Big(\hat L_i-\phi(\hat a^\dagger+\hat a)\Big)^2
+ \hbar\omega_p\!\left(\hat a^\dagger\hat a+\tfrac12\right)
+ \alpha_4\,(\hat a+\hat a^\dagger)^4.
\label{eq:H0}
\end{equation}
We introduce quadratures
$\hat X=\hat a+\hat a^\dagger$, $\hat P=i(\hat a^\dagger-\hat a)$ so $[\hat X,\hat P]=2i$ and define
\begin{equation}
A=\frac{\hbar\omega_p}{4},\quad
B=\frac{\hbar\omega_p}{4}+g\phi^2N,\quad
C=2g\phi,
\end{equation}
as well as $S_2=\sum_{i=1}^N\hat L_i^2$.
Dropping an overall constant, Eq.~\eqref{eq:H0} becomes
\begin{equation}
\hat H_0 = gS_2 + A\hat P^2 + B\hat X^2 - C\,M\,\hat X + \alpha_4\,\hat X^4 .
\label{eq:HPX}
\end{equation}

\paragraph*{Exact sector reduction.}
We proceed by making the displacement $\hat X=x_0+\hat X'$, $\hat P=\hat P'$, with $x_0$ chosen to cancel the linear term:
\begin{equation}
4\alpha_4 x_0^3 + 2B x_0 - C\,M=0.
\label{eq:stationary}
\end{equation}
Among the real roots, we choose the one with positive curvature
$V''(x_0)=2B+12\alpha_4 x_0^2>0$.
After the shift,
\begin{align}
\hat H_0 &= gS_2 + V_{\rm eff}(x_0;M)
+ A\hat P'^2 + B_{\rm eff}\hat X'^2
\nonumber\\
&\qquad + \beta_3\,\hat X'^3 + \alpha_4\,\hat X'^4,
\label{eq:displaced}
\end{align}
with the exact coefficients
\begin{equation}
B_{\rm eff}=B+6\alpha_4 x_0^2,\quad
\beta_3=4\alpha_4 x_0,
\label{eq:coeffs}
\end{equation}
and the sector offset
\begin{equation}
V_{\rm eff}= Bx_0^2 - C M x_0 + \alpha_4 x_0^4.
\label{eq:Veff}
\end{equation}
\begin{comment}
\paragraph*{Closed form for the displacement.}
Equation~\eqref{eq:stationary} is the depressed cubic $y^3+p y+q=0$ with
\begin{equation}
p=\frac{B}{2\alpha_4},\qquad q=-\frac{C\,M}{4\alpha_4}.
\end{equation}
Define the discriminant $\Delta=(q/2)^2+(p/3)^3$. One real root is
\begin{equation}
x_0 = \sqrt[3]{-\frac{q}{2}+\sqrt{\Delta}}
      + \sqrt[3]{-\frac{q}{2}-\sqrt{\Delta}},
\label{eq:x0cardano}
\end{equation}
and when $\Delta<0$ the three real roots follow from the trigonometric form; pick the one with $B_{\rm eff}>0$.

\textcolor{red}{I don't understand anything here. How is (38) real if $p$ is positive? How can the discriminant be negative? And if it can (how?) how would all three roots become real if there is $\sqrt{\Delta}$? And how is root (38) related to those three real roots? There must be three roots in total! And how can $B_{eff}$ ever be negative? }\\
\end{comment}
\paragraph*{Quadratic (Gaussian) spectrum.}
Ignoring the residual anharmonic terms in Eq.~\eqref{eq:displaced} yields a harmonic oscillator with spacing
\begin{equation}
\hbar\Omega(M) = 4\sqrt{A B_{\rm eff}}
= \sqrt{\,\hbar\omega_p\big(\hbar\omega_p+4g\phi^2N+24\alpha_4 x_0^2\big)}.
\label{eq:Omega}
\end{equation}
Hence, in each $(S_2,M)$ sector,
\begin{equation}
E_n^{\rm(G)}(M) = gS_2 + V_{\rm eff}(x_0;M)
+ \hbar\Omega(M)\Big(n+\tfrac12\Big),
%\label{eq:spectrumG}
\nonumber 
\end{equation}
with $n=0,1,2,\dots$.

Here the effective frequency $\Omega$ of the resonator depends nontrivially on the electron configuration $L_{tot}$, as shown in Fig.~3. It can be, for example, directly controlled by resonant radiation.  

For $M=0$ one has $x_0=0$, hence $\beta_3=0$ and 
$\hat h = A\hat P^2 + B\hat X^2 + \alpha_4\hat X^4$ is parity even; $\Omega=\sqrt{\hbar\omega_p(\hbar\omega_p+4g\phi^2N)}$ follows from Eq.~\eqref{eq:Omega}, which is the formula (6) from the main text for the linear cavity. 
\paragraph*{Residual anharmonicity (nonperturbative form).}
The exact single-mode problem in a fixed sector is
\begin{equation}
\hat h \equiv A\hat P'^2 + B_{\rm eff}\hat X'^2 + \beta_3\hat X'^3 + \alpha_4\hat X'^4.
%\label{eq:singlemode}
\nonumber 
\end{equation}
Its spectrum $\{\varepsilon_n\}$ is obtained by numerical diagonalization in a harmonic basis. The full spectrum is then
\begin{equation}
E_n(M) = gS_2 + V_{\rm eff}(x_0;M) + \varepsilon_n(M).
%\label{eq:fullspec}
\nonumber
\end{equation}

\begin{figure}
    \centering
    \includegraphics[width=0.8\linewidth]{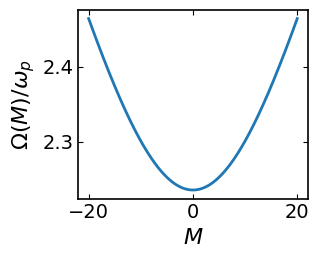}
    \caption{Oscillator frequency $\Omega(M)$ normalized by $\omega_p$ as a function of the total orbital moment $M$ for the anharmonic LC mode. The frequency $\Omega(M) = 4\sqrt{A B_{\mathrm{eff}}(M)}$ is obtained from the curvature of the displaced quartic potential around its $M$-dependent minimum $x_0(M)$. Parameters (in units where $\hbar\omega_p = 1$) are $A = \hbar\omega_p/4$, $B = \hbar\omega_p/4 + g\phi^2 N$, $C = 2g\phi$, with $g = 0.2$, $\phi = 0.5$, $N = 20$, and $\alpha_4 = 0.02$. The mode is softest near $M = 0$ and stiffens symmetrically as $|M|$ increases due to the quartic nonlinearity.
}
    \label{fig:fig3}
\end{figure}

%%%%%%%%%%%%%%%%%%%%%%%%%%
\section{Synthetic Josephson junction in tight-binding systems}

%%%%%%%%%%%%%%%%%%%%

Here we outline the procedure for diagonalizing the  Hamiltonian (29) from the main text, which reads 
\begin{equation}
\hat H_{\{n_k\}}
= -2t\!\left[C\cos(\eta\hat X)-S\sin(\eta\hat X)\right]
+\hbar\omega\,\hat a^\dagger\hat a. 
\label{eq:HCS}
\end{equation}
For numerical diagonalization at arbitrary $\eta$, we work in the oscillator basis $\{\ket{m}\}_{m\ge0}$ where
\begin{equation}
\hbar\omega\,\hat a^\dagger\hat a=\hbar\omega\sum_m m\ket{m}\!\bra{m}.
\end{equation}
Using
\begin{align}
\cos(\eta\hat X)=\tfrac12\!\left(e^{i\lambda(\hat a+\hat a^\dagger)}+e^{-i\lambda(\hat a+\hat a^\dagger)}\right), \nonumber 
\\ 
\sin(\eta\hat X)=\tfrac{1}{2i}\!\left(e^{i\lambda(\hat a+\hat a^\dagger)}-e^{-i\lambda(\hat a+\hat a^\dagger)}\right),
\nonumber 
\end{align}
with $\lambda=\eta/\sqrt2$, the matrix elements are found in closed form:
\begin{align}
& \bra{m}e^{\pm i\lambda(\hat a+\hat a^\dagger)}\ket{n}
= \nonumber \\
& e^{-\lambda^2/2}
\sqrt{\frac{\min(m,n)!}{\max(m,n)!}}\,
(\pm i\lambda)^{|m-n|}\,
L^{(|m-n|)}_{\min(m,n)}(\lambda^2),
\nonumber 
%\label{eq:LaguerreME}
\end{align}
where $L^{(\alpha)}_\ell$ are generalized Laguerre polynomials.
Thus, for each fixed $(C,S)$ one obtains a finite truncated Hermitian matrix for Eq.~\eqref{eq:HCS} converging rapidly in the cutoff.

Equivalently, in the $\hat X$ representation one solves the 1D Schr\"odinger equation
\begin{align}
& \Bigg\{
\frac{\hbar\omega}{2}\!\left(-\frac{d^2}{dX^2}+X^2\right)
-2t\left[C\cos(\eta X)-S\sin(\eta X)\right]
\Bigg\}\psi(X)  \nonumber \\
& = E\,\psi(X),
\nonumber 
%\label{eq:Xrep}
\end{align}
which is exact within the chosen fermionic sector.

\bibliographystyle{apsrev4-2}
\bibliography{References}

\end{document}